\documentclass[twocolumn,showpacs,preprintnumbers,prl,nofootinbib]{revtex4-1}
\usepackage{graphicx,color}

\usepackage{amssymb}
\usepackage{amsbsy}
\usepackage{amsthm,mathrsfs,amsopn}
\usepackage{amsmath}

\begin{document}

\title{Turing instabilities from a limit cycle}

\author{Joseph D. Challenger$^{1,2}$, Raffaella Burioni$^{3}$, Duccio Fanelli$^{2}$}
\affiliation{
1. Department of Infectious Disease Epidemiology, Imperial College London, London, W2 1PG, UK
2. Dipartimento di Fisica e Astronomia, Universit\`{a} di Firenze, INFN and CSDC, Via Sansone 1, 50019 Sesto Fiorentino, Firenze, Italy \\
3. Dipartimento di  Fisica e Scienza della Terra and INFN, Universit\`{a} di Parma, viale G. P. Usberti 7/A 43124, Parma, Italy  }

\begin{abstract} 
The Turing instability is a paradigmatic route to patterns formation in reaction-diffusion systems. Following a diffusion-driven instability, 
homogeneous fixed points can become unstable when subject to external perturbation. As a consequence, the system evolves towards a stationary, nonhomogeneous attractor. Stable patterns can be also obtained via oscillation quenching of an initially synchronous state of diffusively coupled oscillators. In the literature this is known as the oscillation death phenomenon. Here we show that
oscillation death is nothing but a Turing instability for the first return map associated to the excitable system in its synchronous periodic state. In particular we obtain a set of closed conditions for identifying the domain in the parameters space that yields the instability. This is a natural generalisation of the original Turing relations, to the case where the homogeneous solution of the examined system is a periodic function of time. The obtained framework applies to systems embedded in continuum space, as well as those defined on a network-like support. The predictive ability of the theory is tested numerically, using different reaction schemes. 
\end{abstract}

\maketitle

\section{Introduction}

From chemistry to physics, passing through biology and ecology, patterns are widespread in nature. 
Under specific conditions, the spontaneous drive to self-organisation which acts on an ensemble of 
interacting constituents materialises in a rich zoology of beautiful motifs, that bear  
intriguing universal traits \cite{mimura, maron, baurmann, riet, meinhardt, harris, maini, bhat, miura, zhab}. 
The spirals that originate from chemical reactions, 
the stripes in fish skin patterning, the feline coat coloration and the spatial patterns in dryland vegetation 
are all examples of the intrinsic ability of seemingly different systems to yield 
regular structures, both in space and time. 

In 1952 Alan Turing wrote a seminal paper \cite{turing} on the theory of morphogenesis, establishing the mathematical principles 
that drive the process of pattern formation. To this end he considered the coupled evolution of two spatially distributed species, 
subject to microscopic reactions and freely diffusing in the embedding medium. Working in this context, Turing proved that an homogeneous 
mean-field solution of the examined reaction diffusion system can be 
unstable to external perturbations. The Turing instability, as the effect is nowadays called, is seeded by 
diffusion and requires an activator-inhibitor scheme of interaction between agents. When the 
condition for the instability are met, the perturbation grows exponentially in the linear regime. The system subsequently evolves 
towards an asymptotic stationary stable solution characterised by a patchy, spatially inhomogeneous, 
density distribution, which indirectly reflects the collection of modes made unstable at short time and the geometry of the 
hosting support \cite{murray}. Travelling waves can also set in following a symmetry breaking instability of a homogeneous fixed point. 

Turing instabilities are classically studied on regular lattices or continuous supports. For a large class of problems, however, the 
inspected system is defined on a complex network. The theory of patterns formation extends to this latter case, as 
discussed in the pioneering paper by Othmer and Scriven \cite{othmer}, and recently revisited by Nakao and Mikhailov \cite{nakao}. Reaction-diffusion systems defined on a graph can produce an effective segregation into activator-rich and activator-poor nodes, Turing-like 
patterns on a heterogeneous spatial support.    

In the classical Turing paradigm, the conditions for the onset of 
the instability are derived via a linear stability analysis, which requires expanding the imposed perturbation on the complete basis formed 
by the eigenvectors of the (continuum or discrete) Laplacian \cite{nakao,asllaniNature}.  Compact inequalities, containing the entries of the Jacobian matrix for the linearised problem and the diffusion constants for the interacting species, are then derived which constitute the necessary condition for the instability to develop \cite{murray}. 

The formation of a nonuniform stationary state has also been observed in the dynamics of diffusively coupled oscillators. Weak coupling of non-linear oscillators leads to synchronisation, a fundamental phenomenon in nonlinear dynamics which plays a pivotal role in many 
branches of science. Oscillation quenching is an interesting related phenomenon, which is seen in spatially coupled systems \cite{koseka}. Indeed, the possibility of disrupting the oscillations could be in principle exploited as an efficient dynamical regulator \cite{kim_control,kumar_control}. Moreover, it could be implicated in pathological neuronal 
derive, as in the Alzheimer and Parkinson disease. Two different types of oscillation quenching phenomena are generally distinguished in the 
literature, which differ both in the fundamental mechanisms of generation, as well as in their respective manifestations. The suppression of the oscillations can yield a 
final homogeneous steady state, a dynamical process that is known as amplitude death. Oscillation death (OD) is instead 
observed when the initially synchronised state evolves towards an asymptotic inhomogeneous steady configuration \cite{zhou1,zhou2,nakao1}, in response to an externally imposed 
perturbation \cite{koseka}. As remarked upon in the literature (see e.g. Ref.~\cite{zhou1}), the OD pathway is reminiscent of the Turing symmetry-breaking transition, which, as we here recall, originally assumes a reaction-diffusion system perturbed 
around an homogeneous, time-independent, equilibrium. 

Models exhibiting amplitude or oscillation death are, however, difficult to investigate. To progress in the analysis it is customary to invoke a normal 
form representation for the amplitude of the unstable modes near a Hopf bifurcation. Less attention has been devoted to inspecting multispecies reaction-diffusion systems, for which the analysis proves more cumbersome. Alternatively, the master stability formalism \cite{pecora} can be employed to determine the stability (via the largest
Floquet exponent \cite{floquet2}) of the synchronous state, at a given coupling strength.  

Building on these concepts, the aim of this paper is to shed further light on the analogy between OD and Turing instability and eventually base it on solid, quantitative, grounds. As we shall prove in the following, OD is nothing but a Turing instability for the first return map associated to the excitable system in its synchronous periodic state. Arguing along these lines, we will obtain a set of closed conditions for identifying the domain in the parameters space the yields the sought instability. Such conditions constitute an obvious generalisation of Turing original relations, to the interesting setting where the homogeneous solution of the examined system is a periodic function of time. The usual Turing inequalities are recovered when the limit cycle collapses to a fixed point, thus revealing a generalised picture which is consistent with the classical paradigm for pattern formation. The obtained framework holds both for systems embedded in continuum space, as well as for those defined on a network-like support.  The predictive ability of the theory will be demonstrated for different reaction schemes. 

The paper is organised as follows. In the next section we shall review the fundamentals of the Turing instability theory. Then we will move on to studying the effect of a tiny 
heterogeneous perturbation acting on a collection of synchronous reaction-diffusion oscillators. To this end we will make use of the master stability function approach, 
complemented by standard Floquet analysis. Then, in Section III we will derive the generalised Turing conditions to which we alluded above, and test their accuracy 
versus the Floquet-based scenario, for the Brusselator and the Schnakenberg models. Numerical simulations are also reported to illustrate the characteristics of the 
patterns that are asymptotically attained by the systems. To this end we shall consider the systems defined on a two dimensional continuum domain, subject to periodic boundary conditions,  as well as on a heterogeneous network of the Watts-Strogatz type.  Finally in Section V we sum up and draw our conclusions.

\section{Basic Theory of the Turing instability}
\label{TuringMethod}

Consider the following reaction-diffusion system

\begin{eqnarray}
\frac{\partial \phi}{\partial t} &=& f(\phi,\psi)+ D_{\phi} \nabla^2 \phi \nonumber\\
\frac{\partial \psi}{\partial t} &=& g(\phi,\psi)+D_{\psi} \nabla^2 \phi,
\label{eq:reac_dif}
\end{eqnarray}
where $\phi(\mathbf{r},t)$ and $\psi(\mathbf{r},t)$ denote the concentration of the interacting species of respective diffusion constants  $D_{\phi}$ and $ D_{\psi}$. The 
position in space is specified by the vector $\mathbf{r}$ and $t$ stands for time; $f(\cdot,\cdot)$ and  $g(\cdot,\cdot)$ are non-linear functions of the concentrations and represent the reaction contributions. We assume that a stable homogeneous fixed point exists, so that $\phi(\mathbf{r})=\bar{\phi}$ and  
$\psi(\mathbf{r})=\bar{\psi}$, with $\bar{\phi}$ and $\bar{\psi}$ constants, such that $f(\bar{\phi},\bar{\psi})=g(\bar{\phi},\bar{\psi})=0$. To formally verify the stability of the fixed point we introduce the Jacobian matrix $\textbf{J}$:
\begin{equation} 
 \textbf{J}= \left( \begin{matrix} f_{\phi} & f_{\psi}\\ g_\phi& g_{\psi} \end{matrix} \right).
\end{equation}
Here $f_{\phi}$ stands for the derivative of $f$ with respect to the density $\phi$ evaluated at the fixed point 
 $(\bar{\phi},\bar{\psi})$. Similar considerations hold for  $f_{\psi},  g_{\phi},  g_{\psi}$. The homogeneous fixed point is stable provided that
\begin{eqnarray}
\label{hom_stability}
\textrm{tr}(\textbf{J}) &=& f_{\phi} + g_{\psi} <0  \\
\textrm{det}(\textbf{J}) &=& f_{\phi} g_{\psi}- f_{\psi} g_{\phi} >0,
\end{eqnarray}
where $\textrm{tr}(\cdot)$ and $\textrm{det}(\cdot)$ denote respectively the trace and the determinant. The Turing idea consists of introducing a small perturbation 
 $\textbf{w}$ of the initial homogeneous stationary state and looking for the conditions that eventually yield to the growth of such disturbance. In formulae, we set:
\begin{equation} 
 \textbf{w}= \left(\begin{matrix} \delta \phi \\ \delta \psi \end{matrix} \right) \equiv 
\left( \begin{matrix} \phi-\bar{\phi} \\  \psi-\bar{\psi} \end{matrix} \right).
\end{equation}
By hypothesis $|\textbf{w}|$ is small, so we can linearise system (\ref{eq:reac_dif}) around the fixed point to eventually obtain:
\begin{equation} 
\label{linearw}
 \dot{\textbf{w}}=\textbf{J} \textbf{w} + \textbf{D} \nabla^2 \textbf{w},
\end{equation}
where $\dot{\textbf{w}}$ represents the time derivative of $\textbf{w}$ and $\textbf{D}$ is the diagonal diffusion matrix:
\begin{equation} 
 \textbf{D}= \left(\begin{matrix} D_{\phi} & 0 \\ 0 & D_{\psi} \ \end{matrix} \right).
\end{equation}
To solve the above system subject to specific boundary conditions one can introduce the eigenfunctions $\textbf{W}_k(\mathbf{r})$ of the Laplacian as:
\begin{equation} 
 -\nabla^2  \textbf{W}_k(\mathbf{r}) = k^2 \textbf{W}_k(\mathbf{r}),
\end{equation}
for all $k \in \sigma$, where $\sigma$ is a suitable (unbounded) spectral set. We can the expand the perturbation $\textbf{w}$ as:
\begin{equation} 
\label{cond_det}
\textbf{w}(\mathbf{r},t) =  \sum_{k\in \sigma} c_k e^{\lambda(k) t}  \textbf{W}_k(\mathbf{r}),
\end{equation}
where the constants $c_k$ are determined by the initial condition. This operation is equivalent to performing a Fourier transform in space of the original linearised 
equations. The complex function $\lambda(k)$, also known as the dispersion relation, controls the growth or damping of the initial perturbation. The solution of the linearised system exists provided that
\begin{equation} 
\det \left( \lambda I -  \textbf{J}(k^2)  \right) = 0,
\end{equation}
where $I$ is the $2 \times 2$ identity matrix and  $\textbf{J}(k^2)$ is the modified Jacobian matrix with the inclusion of the spatial components, namely:
\begin{equation} 
\label{Jk} 
 \textbf{J}(k^2)= \left( \begin{matrix} f_{\phi}-D_{\phi} k^2 & f_{\psi}\\ g_\phi& g_{\psi} -D_{\psi} k^2 \end{matrix} \right).
\end{equation}
From Eq. (\ref{cond_det}) one obtains the characteristic polynomial:
\begin{equation}
\lambda^2 - B \lambda + C = 0,
\label{char_polyn}
\end{equation}
where:
\begin{eqnarray}
B &=&  f_\phi  + g_\psi  - \left(D_{\phi} + D_{\psi} \right)k^2 \\ \nonumber
C &=&  D_\phi D_\psi k^4 -   \left(D_\phi g_\psi  + D_\psi  f_{\phi}    \right) k^2 + \\ \nonumber
&+&  f_\phi  g_\psi  - f_\psi  g_\phi.
\label{char_polyn}
\end{eqnarray}
Since we are interested in the growth of the unstable perturbation, we should select the largest $\lambda(k) \equiv \lambda_{max}$ which can be written as
\begin{equation}
\lambda_{max} = \frac{1}{2} \left( B + \sqrt{B^2-4 C} \right).
\label{disprel}
\end{equation}
Recalling that, by hypothesis, $\textrm{tr}(\textbf{J}(0))<0$, one can immediately conclude that $B<0$ for all $k$. Hence,  
the condition of the instability $\lambda_{max}>0$  translates into $C<0$.  To obtain a set of closed analytical conditions 
for the instability, we observe that $C$ is a convex parabola in $k^2$. The minimum of the parabola is located at:
\begin{equation}
k^2_{min} = \frac{\left(D_\phi g_\psi  + D_\psi f_{\phi}   \right)}{2 D_\phi D_\psi},
\label{kmin}
\end{equation}
and the corresponding value of $C$, hereafter called $C^{min}$, reads:
\begin{equation}
C^{min} = -\frac{\left(D_\phi g_\psi  + D_\psi f_{\phi}   \right)^2}{4 \left(D_\phi D_\psi\right)^2}+  f_\phi  g_\psi  - f_\psi   g_\phi.
\label{kmin}
\end{equation}
By imposing $C^{min}<0$ and requiring for consistency reasons $k^2_{min}>0$ yields the following conditions for  the instability to develop:
\begin{eqnarray}
\left( D_\phi  g_\psi  + D_\psi f_{\phi}  \right)^2 &>& 4 D_\phi D_\psi \left( f_\phi  g_\psi  -  f_\psi  g_\phi  \right) \nonumber \\ 
\left( D_\phi  g_\psi  + D_\psi f_{\phi}  \right) &>&0.
\label{tur_cond}
\end{eqnarray}
The above inequalities, complemented with the additional conditions (\ref{hom_stability}), are routinely applied to determine the parameters choice that 
make a reaction-diffusion model unstable to externally imposed perturbation of the homogeneous fixed point. Starting from this point, we shall obtain a 
straightforward generalisation of the classical Turing picture, which includes the oscillation death pathway as one of its possible manifestations.

Before concluding this section we remark that the above analysis can be readily adapted to the case of a system defined on a network of $N$ nodes. 
A concise description of this translation can be found in the Appendix.  

\section{Linear instability analysis around a periodic time-dependent solution: the Floquet approach}

In this section we consider the evolution of an external perturbation on an ensemble of synchronous 
oscillators. Our starting point is again system (\ref{eq:reac_dif}) which we now imagine to admit a homogeneous stable solution $(\bar{\phi}(t),\bar{\psi}(t))$, which is periodic with period $T$.
We therefore require $\bar{\phi}(t+T)= \bar{\phi}(t)$ and  $\bar{\psi}(t+T)= \bar{\psi}(t)$,  
for all time $t$. In general the curve $ \bar{\textbf{x}}(t) \equiv (\bar{\phi}(t),\bar{\psi}(t))$ cannot be calculated in closed form, but it can be determined numerically with a prescribed level of accuracy.  

Before proceeding, one must check the stability of the limit-cycle solution. This fact can be assessed via a direct application of Floquet theory \cite{floquet2}, that we 
will here describe. We begin by focusing on a simplified problem, found by ignoring the spatial components of system (\ref{eq:reac_dif}). In other words, we will commence by studying the uniform counterpart of system (\ref{eq:reac_dif}), where the concentrations are solely dependent on time.

We consider a dynamical path starting close to, but not on, the limit cycle. If the limit cycle is stable the difference between this 
path, here called $\textbf{x}(t)$,  and the limit cycle $\bar{\textbf{x}}(t)$ should decay, as time progresses. Introduce 
$\mathbf{\xi}(t)=\textbf{x}(t)-\bar{\textbf{x}}(t)$, by definition small, and linearise the governing equations to obtain:
\begin{equation}
\label{linear-Floquet}
 \dot{\mathbf{\xi}}=\textbf{J}(t) \mathbf{\xi},
\end{equation}
where the Jacobian matrix is now evaluated at the limit cycle and depends therefore on time. Due to the periodic nature of $\bar{\textbf{x}}(t)$ all elements of 
$\textbf{J}(t)$ are periodic and the Floquet theory is hence applicable. Let us label with $\textbf{X}(t)$ a fundamental matrix of system $(\ref{linear-Floquet})$. Then, for all $t$, there exists a singular, constant matrix $\textbf{B}$ such that \cite{floquet2}:

\begin{equation}
\label{Floquet_FundMatrix}
\textbf{X}(t+T)=\textbf{X}(t)\textbf{B}.
\end{equation}
In addition, the following relation holds:
\begin{equation}
\label{DetFloquet}
\textrm{det} \textbf{B} = \exp \left( \int_0^T tr \textbf{J}(t) \textrm{d}t \right).
\end{equation}
The matrix $B$ depends in general on the choice of the fundamental matrix $\textbf{X}(t)$ employed. Nevertheless, its eigenvalues, and hence determinant, do not. The eigenvalues 
$\rho_1$ and $\rho_2$ of $\textbf{B}$ are usually called the Floquet multipliers of the linearised system (\ref{linear-Floquet}). One can also introduce the corresponding Floquet exponent $\mu_i$ defined via the implicit relation $\rho_i=\exp(\mu_i T)$, for $i=1,2$. If $\rho$ is a characteristic multiplier for (\ref{linear-Floquet}) and 
$\mu$ the associated exponent, one can find a particular solution of (\ref{linear-Floquet}) in the form:
\begin{equation}
\mathbf{\xi}(t) = e^{\mu t} \textbf{p}(t),
\end{equation}
where $\textbf{p}(t)$ is a periodic function of period $T$, i.e. such that $\textbf{p}(t+T)=\textbf{p}(t)$. General solutions of the two dimensional system (\ref{linear-Floquet}) can be therefore cast in the form:
 \begin{equation}
\label{generalsolution}
\mathbf{\xi}(t) = c_1 e^{\mu_1 t} \textbf{p}^{(1)} + c_2 e^{\mu_2 t} \textbf{p}^{(2)},
\end{equation}
where the constants $c_1$ and $c_2$ are determined by the initial conditions. For all linear
expansions about limit cycles arising from first-order
equations, one of the Floquet exponents of the system vanishes ($\mu_1=0$ or, equivalently, $\rho_1=1$) throughout
the limit cycle phase \footnote{The non-linear system being considered admits a periodic solution, the limit cycle, which we called 
$\bar{\textbf{x}}(t)$. One can easily show that $\textrm{d} {\bar{\textbf{x}}(t)}/\textrm{d}t$ is a solution of the linearised problem (\ref{linear-Floquet}). Since 
$\textrm{d}{\bar{\textbf{x}}(t)}/\textrm{d}t$ is also a periodic function of period $T$, then the general solution (\ref{generalsolution}) implies that one of Floquet multipliers, say 
$\rho_1$, must be equal to unity, or, equivalently, $\mu_1=0$.}. 
The remaining exponent $\mu_2$ assumes negative real values. The zero exponent is associated
with perturbations along the longitudinal direction of the
limit cycle: these perturbations are neither amplified
nor damped as the motion progresses. At variance, perturbations
in the transverse direction decays in time if the limit cycle
is stable. Recalling that $\textrm{det} \textbf{B} = \rho_1 \rho_2$,  for a stable limit cycle one has   
$\textrm{det} \textbf{B} = \rho_2 = \exp(\mu_2 T) < 1$ and therefore:
\begin{equation}
\label{FloquetSol}
 \int_0^T \textrm{tr} \textbf{J}(t) \textrm{d}t  =  \int_0^T \left( f_{\phi}(t) + g_{\psi}(t) \right)\textrm{d}t <0,
 \end{equation}
a relation that will prove useful in the following. 

Let us now return to discussing the original problem at hand.  Assume the reaction-diffusion system to be initialised in the region of the parameters 
that yields a stable limit cycle behaviour. Therefore, the concentration depends on time, in a periodic fashion. In addition, we assume a uniform spatial distribution, 
meaning that the oscillators are initially synchronised, with no relative dephasing. We then apply a small, non homogeneous (thus site-dependent) perturbation 
and ask ourselves if the interplay between reaction and diffusion can drive into the system a spontaneous symmetry breaking instability. This is nothing but the oscillation death phenomenon that we here discuss in the framework of a self-consistent reaction diffusion kinetics.  

To answer the question, one can adapt to the scope the Floquet analysis outlined above, considering the generalised linear equation  (\ref{linear-Floquet}), 
with the inclusion of space. More concretely, equation  (\ref{linear-Floquet}) reads:

\begin{equation}
\label{linear-Floquet_gen}
 \dot{\mathbf{\xi}}=\textbf{J}(k^2,t) \mathbf{\xi}.
\end{equation}
The matrix $\textbf{J}(k^2,t)$ is formally given by (\ref{Jk}), and its entries are evaluated at the stable (aspatial) limit cycle $\bar{\textbf{x}}(t)$. Floquet theory 
ensures the existence of a solution of problem (\ref{linear-Floquet_gen}) in the form (\ref{generalsolution}) where now $\mu_1$ and $\mu_2$ depend explicitly on the spatial index $k$. If $\mu_{max}$,  the largest of the $\mu_i$, takes positive values over a bounded window in $k$, the reaction-diffusion system is unstable to the imposed perturbation. 
The latter grows exponentially in time, and progressively disrupts the synchrony of the initial configuration. The largest Floquet exponent $\mu_{max}$ is the 
analogue of the dispersion relation $\lambda_{max}$ for the Turing instability and ultimately sets the route to the phenomenon of oscillation death. Unfortunately, the determination of $\mu_{max}$ follows a purely numerical approach and, at this stage, the similarity between Turing and oscillation death cannot be explored in detail. 

In the next section, we shall discuss an alternative approach to the study of the instability of a perturbed array of  
synchronous oscillators. We will derive clear conditions for the onset of the instability, which 
will allow us to reconcile the Turing paradigm and the oscillation death phenomenon, under a unifying framework. 

\section{Alternative conditions for the diffusion-driven instability of a uniform limit-cycle solution}

We now turn to derive an alternative criterion to identify the region of diffusion driven instability from a uniform limit cycle conditions. Our predictions 
will be then confronted to those obtained following the canonical approach based on the Floquet theory. Let us start from the linearised equation 
(\ref{linear-Floquet_gen}) and imagine to partition the interval $[0,T]$ into a collection of $M$ contiguous sub-intervals $[t_i,t_{i+1}]$. We assume that $M$ is sufficiently large that the width of each sub-interval $\Delta t=t_{i+1}-t_{i}$ can be assumed small. To simplify the reasoning we have assumed a uniform partition, 
but this is not a necessary requirement for the following derivation to hold.  

The idea is to solve the linear equation (\ref{linear-Floquet_gen}) within each (small) window of time duration $\Delta t$, and then use this knowledge to 
estimate the cumulative growth of the perturbation, over one complete loop of the limit cycle. In practical terms, and as already anticipated in the introduction, 
we will look at the stability of the first return map, which is associated to the periodic limit cycle solution of the inspected reaction diffusion kinetics. Inside each 
sub-interval, the perturbation ${\mathbf{\xi}}$ obeys a linear ordinary differential equation with time dependent coefficients.

Such an equation can be approximated using a forward Euler scheme, so to establish a direct link between ${\mathbf{\xi}}_{i+1} = {\mathbf{\xi}}(t_{i+1})$ and ${\mathbf{\xi}}_i = {\mathbf{\xi}}(t_i)$:
\begin{equation} 
\label{euler0}
\xi_{i+1} = \left( I+ \Delta t \textbf{J}(k^2,t_i) \right) \xi_i + O(\Delta t^2).
\end{equation}
To compute the global evolution of the perturbation along the limit cycle, one needs to calculate 
\begin{equation}
\label{euler}
{\mathbf{\xi}}_{M} = \Pi_{j=0}^{M-1} \left[ I+ \Delta t \textbf{J}(k^2,t_j)\right] \xi_0.
\end{equation}
Here one must note that the terms in the product must be `time-ordered', with the earlier times to the right.
Neglecting the terms which scale as $\Delta t^n$ with $n \ge 2$, in agreement with the approximated expression (\ref{euler0}) yields:  
\begin{equation}
\label{euler1}
{\mathbf{\xi}}_{M} \simeq  \left[ I+ \Delta t \sum_j \textbf{J}(k^2,t_j)\right] \xi_0.
\end{equation}
In the limit $\Delta t \rightarrow 0$ (which implies sending simultaneously $M \rightarrow \infty$), one can replace the above sum with an integral and write the 
mapping from $\xi_0$ to $\xi_M$ as:  
\begin{equation}
\label{euler2}
{\mathbf{\xi}}(T) =  \left[ I+ \int_0^T \textbf{J}(k^2,t) dt \right] \xi_0 \simeq  \exp \left( \int_0^T \textbf{J}(k^2,t) dt \right) \xi_0.
\end{equation}
Higher order corrections can be also estimated by replacing the Euler scheme (\ref{euler}) with a refined multi-step approximation of the Runge-Kutta type and performing a similar algebraic manipulation of the equations involved. We leave this extension to future work and present instead a different derivation of the above result, which yields consistent conclusions. 

In fact, a formal solution of equation (\ref{linear-Floquet}) can be written down as:
\begin{equation}
{\mathbf{\xi}}_{i+1} =  \exp[\mathbf{\Omega}(t_{i+1},t_i)] {\mathbf{\xi}}_{i},
\label{lin_prob_sol}
\end{equation}
where  $\mathbf{\Omega}(t_{i+1},t_i)=\sum_{k=1}^{\infty} \mathbf{\Omega}_{k,i}$. The form of the first few  $\mathbf{\Omega}_{k,i}$ elements are:
\begin{eqnarray}
\mathbf{\Omega}_{1,i}&=& \int_{t_i}^{t_{i+1}} \textbf{J}(k^2, \tau_1) \textrm{d} \tau_1 \\ \nonumber
\mathbf{\Omega}_{2,i}&=& \frac{1}{2} \int_{t_i}^{t_{i+1}} \textrm{d} \tau_1 \int_{t_i}^{\tau_1} \textrm{d} \tau_2 \left[\textbf{J}(k^2, \tau_1), \textbf{J}(k^2, \tau_2) \right] \\ \nonumber
\mathbf{\Omega}_{3,i}&=& \frac{1}{6} \int_{t_i}^{t_{i+1}} \textrm{d} \tau_1 \int_{t_i}^{\tau_1} \textrm{d} \tau_2 \int_{t_i}^{\tau_3} \textrm{d} \tau_3  \times \\ \nonumber
&\times& (\left[\textbf{J}(k^2, \tau_1), \left[\textbf{J}(k^2, \tau_2), \textbf{J}(k^2, \tau_3) \right] \right]+ \\ \nonumber
&+&\left[ \left[\textbf{J}(k^2, \tau_3), \textbf{J}(k^2, \tau_2) \right], \textbf{J}(k^2, \tau_1) \right] \nonumber,
\label{coeff_Magnus}
\end{eqnarray}
where $\left[ \cdot, \cdot \right]$ stands for the matrix commutator. The above solution is also known as the Magnus series expansion 
\cite{magnus}. From the definition of the coefficients 
(\ref{coeff_Magnus}), it clearly follows that $\mathbf{\Omega}_{s,i} \simeq O([\Delta t]^s)$. Since, by assumption, $\Delta t$ is small, one can truncate the 
infinite sum in the explicit solution (\ref{lin_prob_sol}). In particular, we will consider explicitly the leading term in the series expansion, to 
quantify the dominant contribution. Upon truncation we have therefore:
\begin{equation}
{\mathbf{\xi}}_{i+1} \simeq \exp \left( \mathbf{\Omega}_{1,i}  \right) {\mathbf{\xi}}_{i}.
\label{lin_prob_sol_approx}
\end{equation}
Making use of the above relation, we can for instance relate $\xi_2$ to $\xi_0$ as::
\begin{equation}
\label{xi2}
{\mathbf{\xi}}_2 = \exp \left( \mathbf{\Omega}_{1,1} \right) {\mathbf{\xi}}_1= 
\exp \left( \mathbf{\Omega}_{1,1} \right) \exp \left( \mathbf{\Omega}_{1,0} \right) {\mathbf{\xi}}_0.
\end{equation}
To progress in the analysis we first recall the Baker-Campbell-Hausdorff formula. Consider two non commuting matrices $\mathbf{Z_1}$ and $\mathbf{Z_2}$. Then, the 
product $\exp(\mathbf{Z_1}) \exp(\mathbf{Z_2})$ can be written as $\exp(\mathbf{Z})$ where:
\begin{equation}
\mathbf{Z}=\mathbf{Z_1}+\mathbf{Z_2}+\frac{1}{2}[\mathbf{Z_1},\mathbf{Z_2}]+ ... ,
\end{equation}
where $[\cdot, \cdot]$ stands for the matrix commutator. If matrices $\mathbf{Z_1}$ and $\mathbf{Z_2}$ commute, namely if $[\mathbf{Z_1},\mathbf{Z_2}]=0$, one recovers the usual formula for the composition of the exponential of scalars. Making use of the above relation in the expression (\ref{xi2}) for ${\mathbf{\xi}}_2$, one obtains:
\begin{equation}
\label{xi2_1}
{\mathbf{\xi}}_2 = \exp \left( \mathbf{\Omega}_{1,1} + \mathbf{\Omega}_{1,0}  + \frac{1}{2} [\mathbf{\Omega}_{1,1},\mathbf{\Omega}_{1,0}] + ... \right) {\mathbf{\xi}}_0.
\end{equation}
The correlator $[\mathbf{\Omega}_{1,1},\mathbf{\Omega}_{1,0}]$ involves the product of terms of order $O(\Delta t^2)$, and it should be therefore neglected for consistency reason, as the expansion is truncated at order $O(\Delta t)$. Moreover, it can be argued that the commutation of matrices defined on neighbours intervals of the 
partition in $t$ scales as $\Delta t^3$, an observation that makes it cumbersome to organise the next to leading corrections in growing powers of $\Delta t$.

The reasoning that we have outlined above can be iterated forward. One gets eventually the  following expression for the magnitude of 
the perturbation ${\mathbf{\xi}}_{M}$, at the considered order of approximation:
\begin{equation}
{\mathbf{\xi}}_M = \exp \left(\sum_{j=0}^{M-1} \Omega_{1j}  \right) {\mathbf{\xi}}_{0}.
\label{lin_prob_sol1}
\end{equation}
Performing the continuum limit ($\Delta t \rightarrow 0$ and $M \rightarrow \infty$) we obtain:
\begin{equation}
{\mathbf{\xi}}(T)  \simeq \exp \left( \int_{0}^{T} \textbf{J}(k^2, \tau_1) \textrm{d} \tau_1  \right) {\mathbf{\xi}}(0) = 
\exp(\langle \textbf{J} \rangle T) {\mathbf{\xi}}_0,
\label{lin_prob_sol2}
\end{equation}
where $\langle \textbf{J} \rangle = (1/T) \int_0^T \textbf{J}(k^2, \tau) \textrm{d} \tau$. The above equation coincides with equation (\ref{euler2}), obtained under the Euler scheme. 

Starting from this setting, it is possible to derive a compact criterion for the onset of the instability, which we will then validate a posteriori 
versus the standard Floquet technique. To this end, we assume $\langle \textbf{J} \rangle$ to be diagonalisable. 
Hence, there exist a matrix  $\textbf{U}$ such that  $\langle \textbf{J} \rangle = \textbf{U}  \textbf{D}_J  \textbf{U}^{-1}$ where $\textbf{D}_J$ is a diagonal matrix. 
Eq. (\ref{lin_prob_sol2}) transforms into:
\begin{equation}
{\mathbf{\xi}}(T) = \exp\left( \textbf{U}  \textbf{D}_J  \textbf{U}^{-1} T \right) {\mathbf{\xi}}(0)= 
\textbf{U}  \exp\left(\textbf{D}_J T \right)  \textbf{U}^{-1} {\mathbf{\xi}}(0).
\label{lin_prob_sol_map3}
\end{equation}
We now introduce $\mathbf{\eta} = \textbf{U}^{-1} \mathbf{\xi}$. The map (\ref{lin_prob_sol2}) 
takes therefore the simple form:
\begin{equation}
\mathbf{\eta}(T) = \exp\left( \textbf{D}_J  T \right) \mathbf{\eta}(0).
\label{lin_prob_sol_map_fin}
\end{equation}
The eigenvalues of the averaged Jacobian matrix $\textbf{J}$ determine the fate of the perturbation. If the real part of the largest eigenvalue is positive, then the perturbation develops, otherwise it fades away after successive iteration of the return map. To derive the condition for the emergence of the instability one must therefore
calculate the eigenvalues $\lambda_{1,2}$ of $\langle \textbf{J} \rangle$, which are the solutions of the following characteristic polynomial:
\begin{equation}
\lambda^2 - B_{(1)} \lambda + C_{(1)} = 0,
\label{char_polyn}
\end{equation}
where:
\begin{eqnarray}
B_{(1)} &=&  \langle f_\phi \rangle + \langle g_\psi \rangle - \left(D_{\phi} + D_{\psi} \right)k^2 \\ \nonumber
C_{(1)} &=&  D_\phi D_\psi k^4 -   \left(D_\phi \langle g_\psi \rangle + D_\psi \langle f_{\phi} \rangle   \right) k^2 + \\ \nonumber
&+&  \langle f_\phi \rangle \langle g_\psi \rangle -\langle f_\psi \rangle \langle g_\phi \rangle, 
\label{char_polyn}
\end{eqnarray}
where $\langle f_\phi \rangle = (1/T)  \int_0^T f_{\phi} dt$. Similarly for $\langle f_\psi \rangle$, 
$\langle g_\phi \rangle$ and $\langle g_\psi \rangle$. 
Hence, the largest real eigenvalue $\lambda_{max}$ is:
\begin{equation}
\lambda_{max} = \frac{1}{2} \left( B_{(1)} + \sqrt{B_{(1)}^2-4 C_{(1)}} \right) .
\label{char_polyn}
\end{equation}
Recalling that by definition $B_{(1)}<0$ (the limit cycle is stable), the condition of the instability $\lambda_{max}>0$ 
translates into $C_{(1)}<0$. This is nothing but the same condition that it is recovered following the conventional Turing calculation with the only difference that now the time dependent entries of the Jacobian matrix are averaged over one complete loop of the unperturbed limit cycle. To obtain closed analytical condition for the instability, 
one can repeat the steps of the derivation reported in Section~\ref{TuringMethod} to eventually get:
\begin{eqnarray}
\left( D_\phi \langle g_\psi \rangle + D_\psi \langle f_{\phi} \rangle  \right)^2 &>& 4 D_\phi D_\psi \left( \langle f_\phi \rangle \langle g_\psi \rangle - \langle f_\psi \rangle \langle g_\phi \rangle \right) \nonumber \\ 
\left( D_\phi \langle g_\psi \rangle + D_\psi \langle f_{\phi} \rangle  \right) &>&0,
\label{tur_cond}
\end{eqnarray}
which constitute a natural generalisation of the standard Turing recipe. Indeed, the above relations reduce to the Turing conditions when the limit cycle converges to a fixed point. 

\section{Numerical validation}

To test the adequacy of the theory, we shall consider two distinct reaction schemes, 
the Brussellator and the Schnakenberg model. For both systems, we will delimit the portion of the relevant parameters space for which the instability is expected to develop, based on conditions (\ref{tur_cond}). These predictions are compared to those obtained using Floquet analysis. Numerical simulations are also performed to challenge the validity of the proposed theoretical picture.   

\subsection{The Brusselator model}

\begin{figure}[h]
\begin{center}
\hspace*{-1cm}
\includegraphics[scale=0.83]{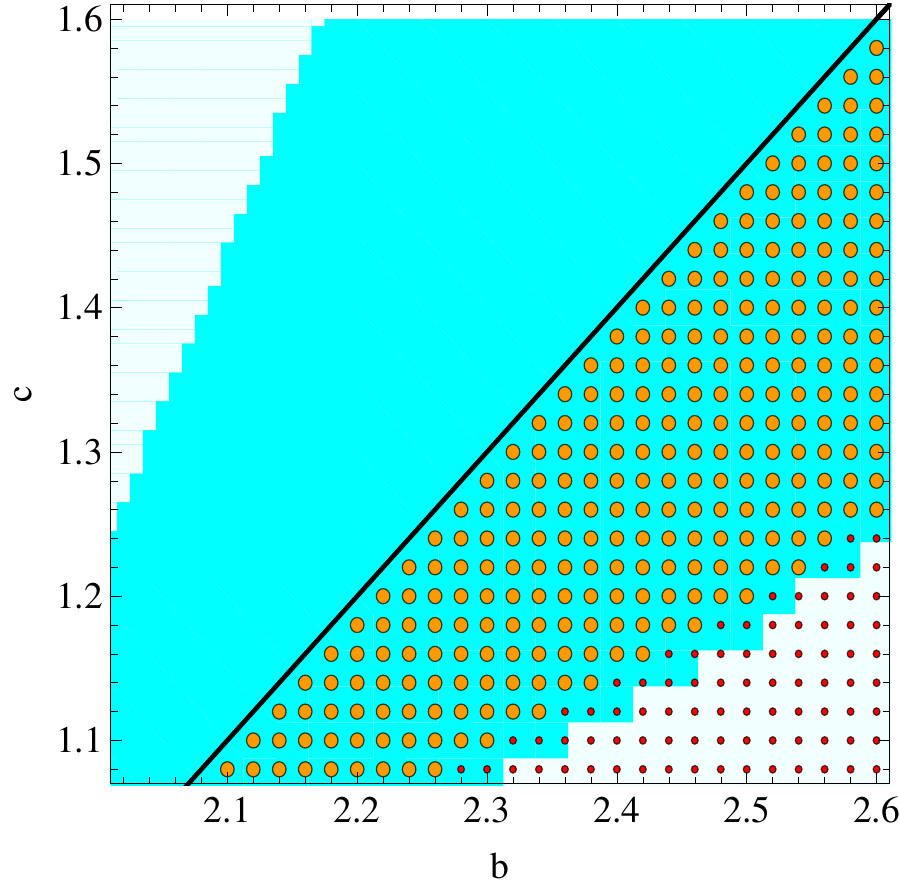}
\end{center}
\caption{The extended region of the Turing instability for the Brusselator model, as parameters $b$ and $c$ are varied. The diffusion coefficients were $D_{\phi}=0.07 $, $D_{\psi}=0.5$. The solid line shows the Hopf bifurcation for the aspatial model: above the line the aspatial
system converges toward a  stable fixed point. Below the line, a stable homogeneous limit cycle solution is instead found. The circular symbols show results from the Floquet approach: the larger symbols indicating an instability, the smaller ones indicating that the homogeneous system 
is stable. The shaded region identifies the region of parameter space where the instability is predicted to occur, following Eq.~\eqref{tur_cond}.
}
\label{fig:Fig1}
\end{figure}

\begin{figure}[h]
\begin{center}
\includegraphics[scale=0.85]{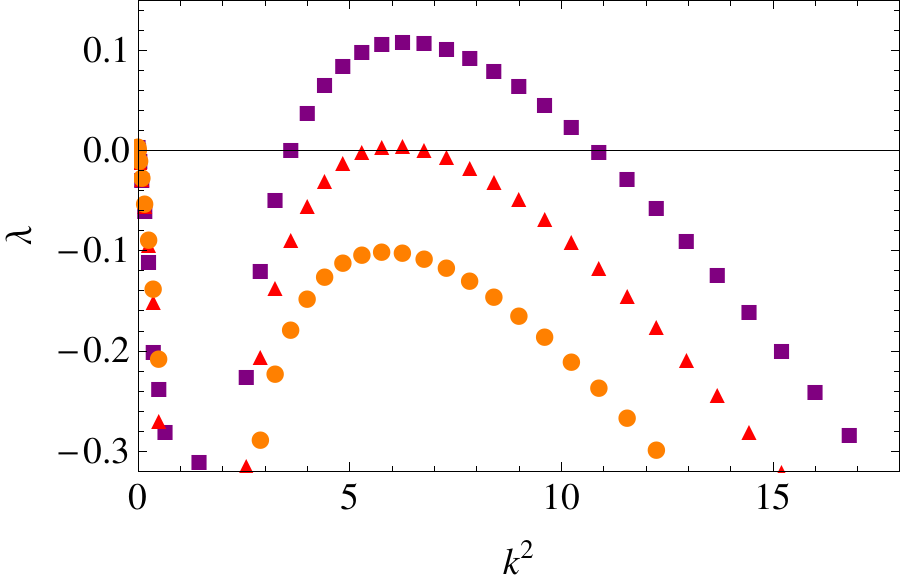}
\end{center}
\caption{Dispersion relations for the Brusselator model for three parameter choices, calculated from the Floquet analysis. Fixing $b=2.5$, we used $c=1.3$ (purple squares), $c=1.2$ (red triangles) and $c=1.1$ (orange circles). The diffusion coefficients were $D_{\phi}=0.07 $, $D_{\psi}=0.5$. }
\label{fig:disp_b}
\end{figure}

Here we will make use of the so-called Brusselator model, a two species reaction-diffusion model whose local reaction terms are $f(\phi,\psi)=1-(b+1) \phi + c \phi^2 \psi $ 
and $g(\phi,\psi)=b \psi - c \phi^2 \psi $, where $b$ and $c$ act as control parameters. Conditions (\ref{tur_cond}) allow us to delimit a compact portion of the parameters plane $(b,c)$ for which the generalised Turing instability is expected to develop. Results of the study are reported in Fig. \ref{fig:Fig1}: 
the solid line separates the fixed point from the limit cycle regime. The region of instability, the shaded area in the figure, extends beyond the Hopf bifurcation, and includes the standard Turing domain as part of it. As discussed earlier, the instability domain inside the region of stable homogeneous limit cycle can be calculated via a direct implementation of the Floquet technique. Large orange circles in Fig. \ref{fig:Fig1} identify the instability domain as computed via the Floquet analysis, while small red dots 
refer to the choice of the parameters for which the OD instability cannot take place. These results 
agree with the prediction obtained from the generalised Turing inequalities (\ref{tur_cond}). In Fig.~\ref{fig:disp_b} we show the dispersion relations for three parameter choices. Fixing $b=2.5$, we vary the value of $c$, showing results from inside and outside the instability region. 

\begin{figure}[h]
\begin{center}
\includegraphics[scale=0.45]{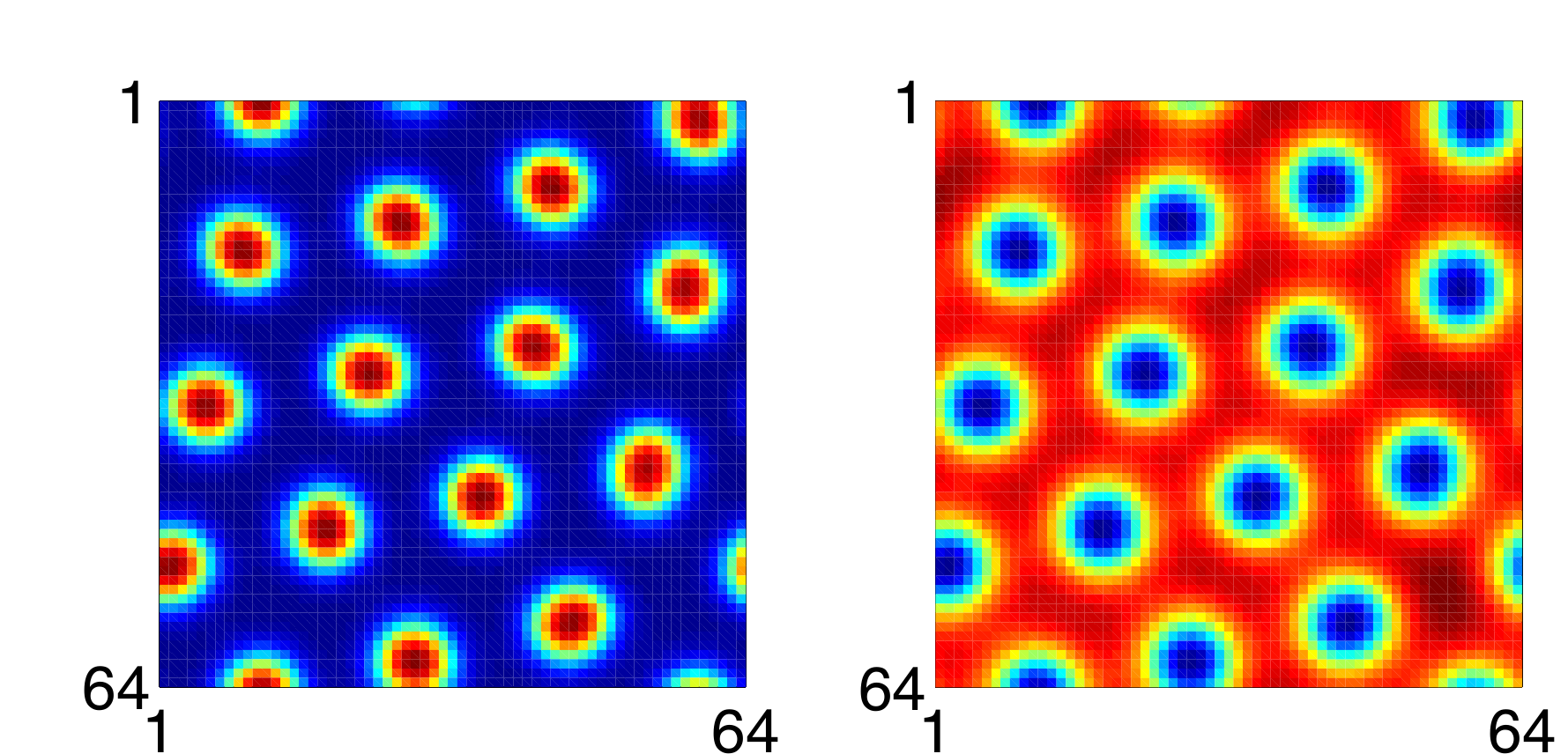}
\end{center}
\caption{The late time evolution for species $\phi$ (left) and $\psi$ (right) for the Brusselator model inside the extended region of Turing-like order. The initial homogeneous limit-cycle state is disturbed by a small non homogeneous perturbation. The synchrony of the spatially coupled oscillators is lost and the system evolves towards a stationary stable configuration. The patterns resemble (indeed, under the Fourier lens, are identical to) the patterns obtained inside the classical Turing region, i.e. above the Hopf transition line. In other words, it looks like the same Turing attractor can be reached following two alternative dynamical pathways.  Parameters are:  $b=2.4 $; $c=1.2 $; $D_{\phi}=0.07 $, $D_{\psi}=0.5 $. The simulations are carried our over a square box of linear size $L=10$ partitioned in $64$ mesh points. }
\label{fig:Fig2}
\end{figure}

\begin{figure}[h]
\begin{center}
\includegraphics[scale=0.35]{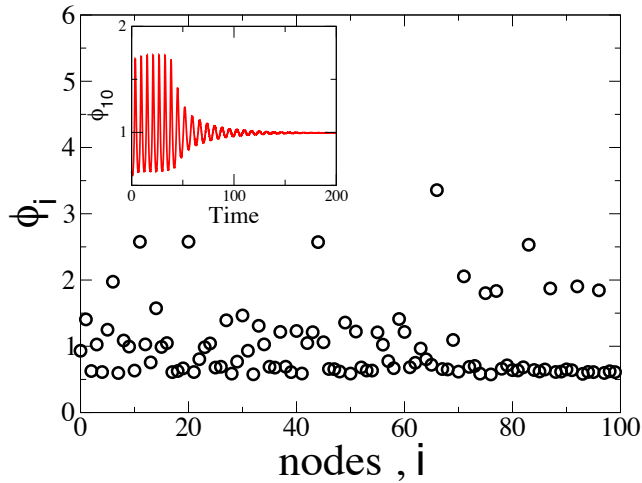}
\end{center}
\caption{Stationary pattern attained by the Brusselator model, defined on a network of the Watts-Strogatz type (number of node $N=100$ and probability of rewiring $p=0.8$). In the main panel, the asymptotic concentration of species $\phi_i$ is plotted as function of the nodes index $i$. In the inset the evolution of the concentration on a particular node is shown, in order to appreciate the transition from the initial oscillatory regime to the final stationary state. The parameters are set as in Fig. \ref{fig:Fig2}.}
\label{fig:Fig3}
\end{figure}

Numerical simulations are also performed for the system initialised inside the extended region of instability to visualise the asymptotic, stationary stable solution that the system eventually attains. To emphasise the broad relevance of our conclusion we performed simulations for (i) the Brusselator model defined on a regular two-dimensional support, subject to periodic boundary conditions (see Fig. \ref{fig:Fig2}); (ii) the Brusselator model defined on a Watts-Strogatz network \cite{watts} (see Fig. \ref{fig:Fig3}).  

\subsection{The Schnakenberg model}

\begin{figure}[h]
\begin{center}
\includegraphics[scale=0.83]{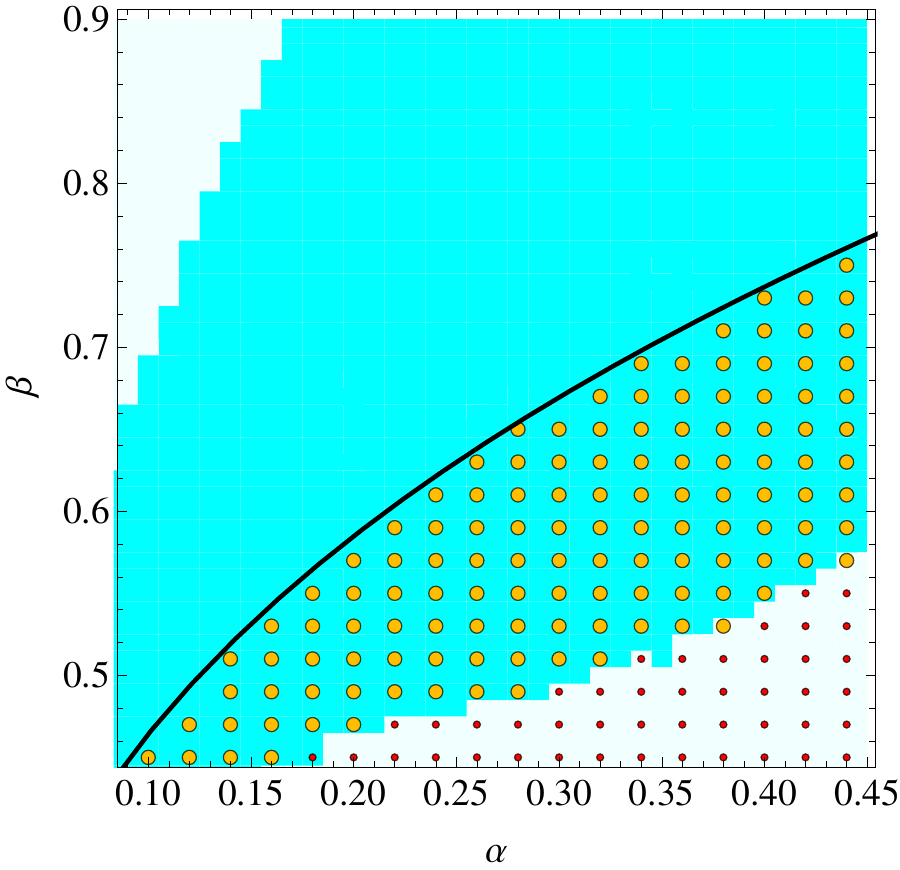}
\end{center}
\caption{
The extended region of the Turing instability for the Schnakenberg model, as parameters $\alpha$ and $\beta$ are varied. The diffusion coefficients were $D_{\phi}=0.01$, $D_{\psi}=1$. The solid line shows the Hopf bifurcation for the aspatial model: above the line the aspatial
system converges toward a  stable fixed point. Below the line, a stable homogeneous limit cycle solution is instead found. The circular symbols show results from the Floquet approach: the larger symbols indicating an instability, the smaller ones indicating that the homogeneous system  is stable. The shading delimits the region where the instability is predicted to occur by  Eqs.~\eqref{tur_cond}. 
}
\label{fig:Fig4}
\end{figure}

\begin{figure}[h]
\begin{center}
\includegraphics[scale=0.85]{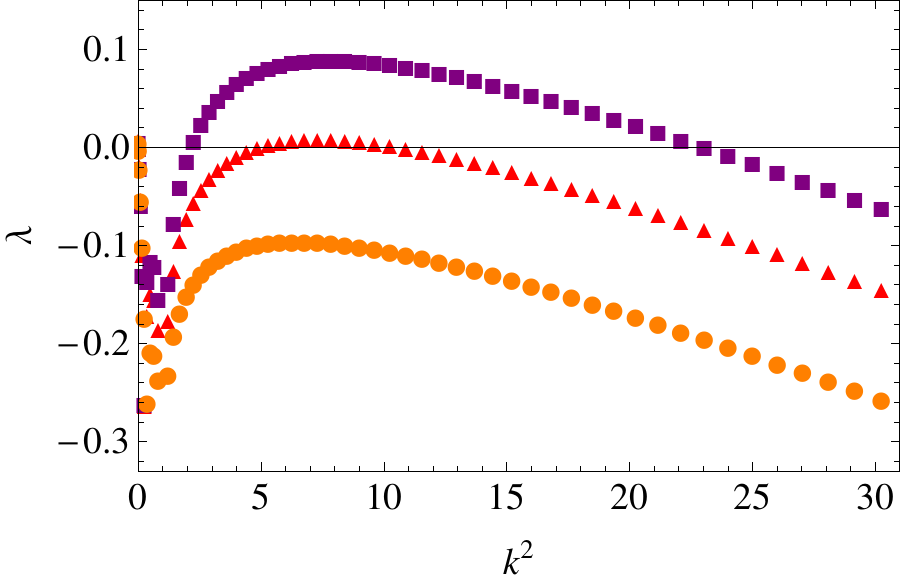}
\end{center}
\caption{Dispersion relations for the Schnakenberg model for three parameter choices, calculated from the Floquet analysis. Fixing $\alpha=0.36$, we used $\beta=0.56$ (purple squares), $\beta=0.52$ (red triangles) and $\beta=0.48$ (orange circles). The diffusion coefficients were $D_{\phi}=0.01 $, $D_{\psi}=1$. }
\label{fig:disp_s}
\end{figure}

\begin{figure}[h]
\begin{center}
\includegraphics[scale=0.45]{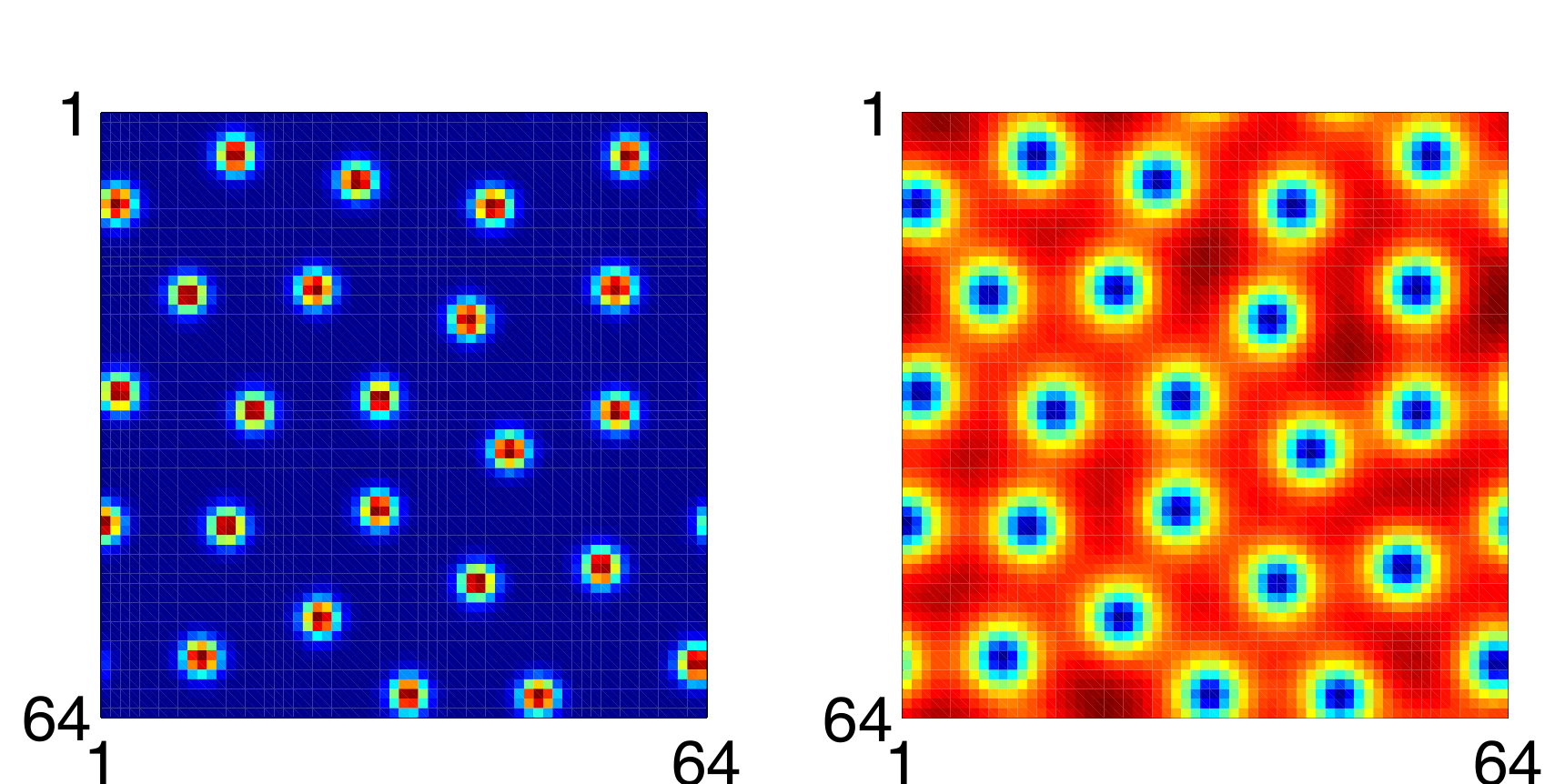}
\end{center}
\caption{The final stationary state obtained for species $\phi$ by initialising the Schnakenberg model inside the region where the homogeneous (hence aspatial) limit-cycle is stable, and imposing a small perturbation to the initially synchronous oscillations. As already remarked in the caption of 
Fig. \ref{fig:Fig2} the patterns are practically indistinguishable from those obtained inside the classical Turing region. Here also, it seems plausible to hypothesise that the same Turing attractor can be reached following alternative dynamical paths. Parameters are $a=0.125$, $b=0.475$ (or $\alpha=0.35$ and $\beta=0.6$), $D_{\phi}=0.01$, $D_{\psi}=1$. The simulations are carried our over a square box of linear size $L=10$ partitioned in $64$ mesh points.   
}
\label{fig:Fig5}
\end{figure}

\begin{figure}[h]
\begin{center}
\includegraphics[scale=0.32]{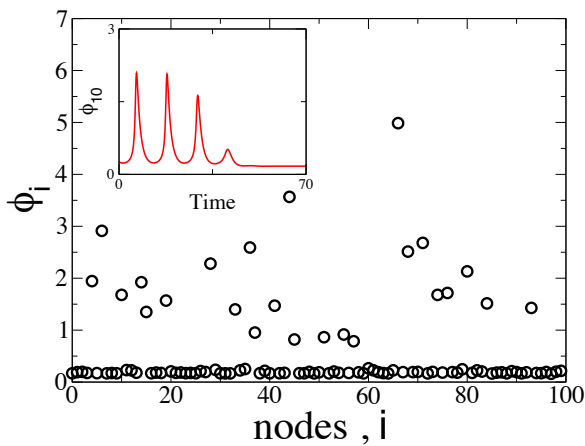}
\end{center}
\caption{Stationary pattern attained by the Schnakenberg model, defined on a network of the Watts-Strogatz type (number of nodes $N=100$ and probability of rewiring $p=0.8$). In the main panel, the asymptotic concentration of species $\phi_i$ is plotted as function of the nodes index $i$. In the inset the time evolution of the concentration on one of the nodes of the network is shown. The transition from the initial oscillation to the final stationary state is clearly displayed. The parameters are set as in Fig. \ref{fig:Fig5}.}
\label{fig:Fig6}
\end{figure}

We shall here consider the Schakenberg model and repeat the analysis reported above. The Schnakenberg model is characterised by 
the following reaction terms: $f(\phi,\psi)=a- \phi + \phi^2 \psi $ and $g(\phi,\psi)=b -  \phi^2 \psi $, where $a$ and $b$ are constant parameters. To study the system it is customary to introduce the parameters $\alpha=b-a$ and $\beta=a+b$.
The shaded area in Fig. \ref{fig:Fig4} identifies the region of the parameters plane $(\alpha, \beta)$ where the instability is predicted to occur. We again emphasise that 
patterns are expected to occur outside the region of classical Turing order, well inside the domain where the aspatial models displays a stable limit cycle solution. As for the case of the Brusselator model, one reaches consistent conclusions if the Floquet analysis is employed instead of  Eqs. (\ref{tur_cond}), the generalised Turing inequalities. In Fig.~\ref{fig:disp_s} we show the dispersion relations for three parameter choices. Fixing $\alpha=1.3$, we vary the value of $\beta$, showing results from inside and outside the instability region. Numerical simulations for the Schnakenberg system defined both on a continuous two dimensional support and on a heterogeneous complex networks are performed and the asymptotic, stationary stable, patterns displayed in Figures \ref{fig:Fig5} and \ref{fig:Fig6}, respectively.  

\section{Conclusion}

Reaction-diffusion systems display a rich plethora of interesting solutions.  Particularly relevant is the 
spontaneous emergence of self-organised stationary patterns, originating from a symmetry breaking instability of a 
homogeneous fixed point. The dynamical mechanism that seeds such instability was illustrated by Alan Turing in his 
pioneering work on the chemical basis of morphogenesis. Since then, it has been exploited in many different contexts, 
ranging from physics to biology.  The concept of the Turing instability also applies to reaction-diffusion systems defined 
on a complex network, a setting that is of paramount importance for neuroscience-related applications. The internet and the cyberworld in general 
are other obvious examples which require the concept of network. 

Beyond the Turing picture, stationary regular motifs can also 
originate from oscillation quenching of a spatially extended chain of coupled non-linear 
oscillators. This phenomenon, usually referred to as oscillation death, has been mainly investigated  
by resorting to a normal form approximation for the evolution of the spatially unstable modes. Mathematical progress is possible via a 
semi-analytical approach which combines knowledge from the celebrated master stability formalism \cite{pecora} to the Floquet technique.

Starting from this setting, we have here investigated the process of pattern formation for a multispecies model, which displays a limit cycle behaviour in 
its aspatial limit. We have showed that oscillation death is nothing but the classical Turing instability  for the first return map associated to the excitable 
system in its synchronous periodic state. Working along these lines we have obtained a system of compact inequalities, which set the conditions for the onset of the 
instability. The obtained conditions constitute a natural generalisation of the Turing recipe, so as to include the case where the imposed perturbation acts on  
a homogeneous time-dependent periodic solution. The proposed criterion returns a wider region of Turing instability, as compared to the conventional approach. 
The stationary patterns that originate from the inhomogeneous perturbation of the limit cycle solution are virtually indistinguishable for those obtain within the classical Turing region, as we demonstrated with reference to specific case studies. Based on these findings, we propose that the conditions for the generalised instability that we have derived should be carefully considered for all reaction-diffusion schemes, which undergo Turing ordering while displaying a limit cycle solution in their aspatial counterpart versions.

\section{Appendix}

The purpose of this Appendix is to briefly discuss the generalisation of the above analysis to the relevant setting where the reaction diffusion system is defined on a 
discrete support, such as a complex heterogeneous network.

We begin by considering a network made of $N$ nodes and characterised by the $N \times N$ adjacency matrix $W$: the entry $W_{ij}$ is equal to one if nodes $i$ and $j$ 
(with $i \ne j$) are connected, and it is zero otherwise. If the network is undirected, the matrix $W$ is symmetric.  
A general reaction-diffusion system defined on the network reads:
\begin{eqnarray}
\frac{d \phi_i}{d t} &=& f(\phi_i,\psi_i)+ D_{\phi} \sum_j \Delta_{ij} \phi_j \nonumber\\
\frac{d \psi_i}{d t} &=& g(\phi_i,\psi_i)+D_{\psi} \sum_j \Delta_{ij} \psi_j.
\label{eq:reac_dif-network}
\end{eqnarray}
Here $\Delta_{ij}=W_{ij}-k_i \delta_{ij}$ is the network Laplacian,  $k_i$ stands for the connectivity of 
node $i$ and $\delta_{ij}$ is the Kronecker's delta. Assume now that a homogeneous fixed point of system 
(\ref{eq:reac_dif-network}) exists and indicate it with $(\bar{\phi},\bar{\psi})$. The fixed point is stable provided 
eqs.~(\ref{hom_stability}) hold. Patterns arise when  $(\bar{\phi},\bar{\psi})$ becomes unstable to inhomogeneous 
perturbations. As already discussed with reference to the continuum setting, one can introduce a small perturbation
$(\delta \phi_i, \delta \psi_i)$ to the fixed point and linearise around it, to look for the conditions that seed the instability. 
One obtains a linear equation which is equivalent to eq. (\ref{linearw}) except for the index $i$ which is attached to the perturbation amount, and hence to 
$\textbf{w}$, and which reflects the discreteness of the embedding structure. To solve the linear problem one needs to introduce the spectrum of the 
Laplacian operator:
\begin{equation}
\sum_{j=1}^N \Delta_{ij} v_j^{(\alpha)} = \Lambda^{(\alpha)} v_i^{(\alpha)}, \qquad \alpha=1,...,N,
\end{equation} 
where $\Lambda^{(\alpha)}$ and $v_i^{(\alpha)}$ respectively represent the eigenvalues and their associated eigenvectors. Then,
the inhomogeneous perturbation can be expanded as:
\begin{eqnarray}  
\delta \phi_i &=& \sum_{j=1}^N c_{\alpha} e^{\lambda_{\alpha}t} v_i^{(\alpha)} \\
\delta \psi_i &=& \sum_{j=1}^N b_{\alpha} e^{\lambda_{\alpha}t} v_i^{(\alpha)},
\end{eqnarray}
where the constants $c_{\alpha}$ and $b_{\alpha}$ refer to the initial condition. By inserting the above expression in the equation which governs the evolution of the perturbation at the linear order, one gets a dispersion relation which is identical to (\ref{disprel}), provided the factor $-k^2$ is replaced with the Laplacian eigenvalues
$\Lambda^{(\alpha)}$. In practice, it is this latter quantity which determines the spatial characteristic of the emerging patterns, when the system is defined on a heterogeneous complex support. Obviously, inequalities (\ref{tur_cond}) extend to the case of networks, noting that $-k^2$ hands over into $\Lambda^{(\alpha)}$. The discussion above adapts easily to the case where the perturbation is studied around a homogeneous limit cycles solution.

\end{document}